\documentclass[journal]{IEEEtran}
\usepackage{set space}

\usepackage{cite}

\ifCLASSINFOpdf
   \usepackage[pdftex]{graphicx}
   \graphicspath{./}
  \DeclareGraphicsExtensions{.pdf,.jpeg,.png}
\else
  \usepackage[dvips]{graphicx}
\fi
\usepackage{subfigure}

\usepackage{color}

\usepackage{amsmath}
\usepackage{amsfonts,helvet}
\usepackage{amssymb}

\usepackage{algorithm}
\usepackage{algpseudocode}

\usepackage{array}
\usepackage{multirow}

\begin{document}

\title{Large-scale Antenna Operation in Heterogeneous Cloud Radio Access Networks: A Partial Centralization Approach
}

\author{Sangkyu~Park,~\IEEEmembership{Student Member,~IEEE,}
        Chan-Byoung~Chae,~\IEEEmembership{Senior Member,~IEEE,} and\\
        Saewoong~Bahk,~\IEEEmembership{Senior Member,~IEEE}
}


%


\maketitle

\begin{abstract}

 To satisfy the ever-increasing capacity demand and quality of service (QoS) requirements of users, 5G cellular systems will take the form of heterogeneous networks (HetNets) that consist of macro cells and small cells. To build and operate such systems, mobile operators have given significant attention to cloud radio access networks (C-RANs) due to  their beneficial features of performance optimization and cost effectiveness. Along with the architectural enhancement of C-RAN, large-scale antennas (a.k.a. massive MIMO) at cell sites contribute greatly to increased network capacity either with higher spectral efficiency or through permitting many users at once. In this article, we discuss the challenging issues of C-RAN based HetNets (H-CRAN), especially with respect to large-scale antenna operation. We provide an overview of existing C-RAN architectures in terms of large-scale antenna operation and promote a partially centralized approach. This approach reduces, remarkably, fronthaul overheads in C-RANs with large-scale antennas. We also provide some insights into its potential and applicability in the fronthaul bandwidth-limited H-CRAN with large-scale antennas.

\end{abstract}

\begin{IEEEkeywords}
C-RAN, cloud radio access network, massive MIMO, large-scale antenna system, fronthaul.
\end{IEEEkeywords}

%
\IEEEpeerreviewmaketitle

\section{Introduction}
 
From 2010 to 2015, mobile data traffic has grown, experts believe, more than 24 times; from 2010 to 2020, they expect it to grow more than 500 times \cite{Cisco_TR_2013}.
To satisfy the explosive capacity demand, mobile operators need to increase network capacity by adding more cells, thereby creating a complicated structure of heterogeneous and small cell networks (HetSNets).

To implement HetSNets, mobile operators are closely looking  at a cost effective and performance-optimizing radio access architecture known as cloud radio access network (C-RAN)\cite{CMRI}. In conventional RANs, Layer-2 and PHY-layer functions are processed at antenna sites while in C-RANs, most base station functionalities are processed at central base band unit (BBU) pools, and the remaining, minimal radio frequency functionalities are processed at remote radio heads (RRHs). 

Owing to the centralization of computing resources and radio signal generation functions, equipment at cell-sites can be cheaper and computing resources more efficiently utilized in a cloud system. As the baseband radio signals are manufactured at the central BBU pool, it becomes easy to upgrade the radio access technology (RAT) and support multi-RAT. For mobile operators, these modifications result in reduced capital expenditure (CAPEX) and operating expenditure (OPEX). 
 
Moreover, clustering of cross-tier and co-tier cells makes it easy to optimize network-wide performance by exploiting, in real-time, inter-cell coordination techniques such as Coordinated Multi-Point (CoMP), joint spectrum resource allocation, and scheduling as well as multiple associations.

\begin{figure*}
  \centering
  \includegraphics[width=7in]{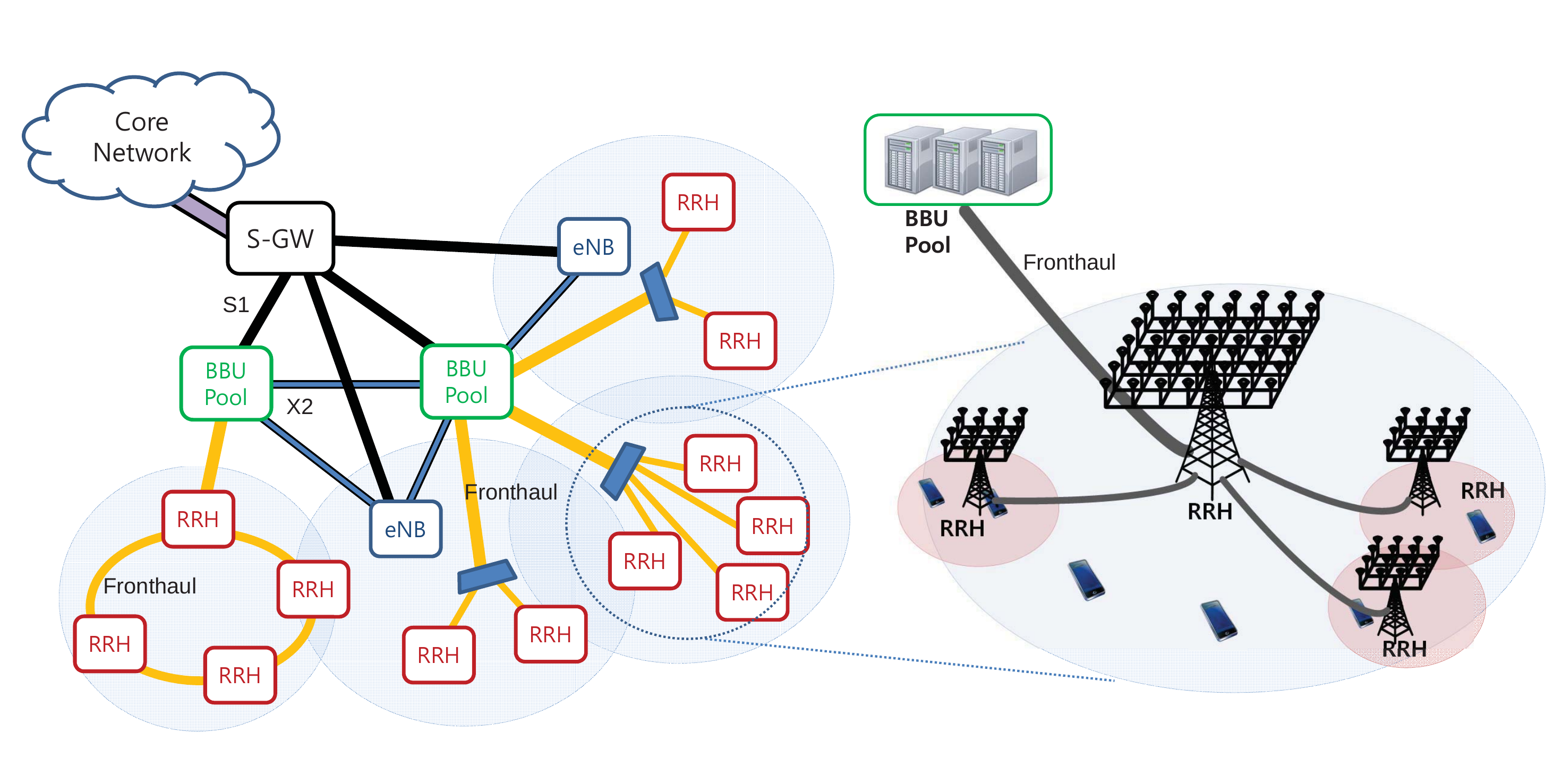}
  \caption{H-CRAN architecture and deployment scenario.} \label{fig:Fig_Scenario_HCRAN}
  \end{figure*}

Due to strict requirements on capacity and delay that stimulate such centralized resource management and baseband processing, underlay transport networks among BBU pools and RRHs are commonly constructed via optical networks. Fig. \ref{fig:Fig_Scenario_HCRAN} shows a HetSNet C-RAN (H-CRAN) and fronthaul transport network scenario where multiple RRHs for macro and small cells are single- or multi-hop connected to central locations (i.e., BBU pools), sharing fiber links.

In boosting per-cell capacity, a key enabler is the large-scale antenna systems (a.k.a. massive MIMO), in parallel. A large array of antenna elements enable narrow beamforming for many simultaneous users with a large degree of freedom. If detailed channel state information (CSI) is available, a  base station can simultaneously serve tens-or hundreds of terminals (but fewer than the number of antennas) by exploiting proper transmit precoding or receive combining with the large set of antennas.

The challenge arises when large-scale antenna systems must be implemented with H-CRAN. The centralization concept that seems absolutely beneficial precipitates, as the number of antennas increases at cell sites, explosive data volume into fronthaul links. 
Therefore, to support increasing fronthaul data, mobile operators are forced to add more fiber cables and optical devices.

In this article, we provide an overview of the challenging issues of H-CRAN with large-scale antennas and investigate C-RAN architectures in terms of large-scale antenna operation in H-CRANs. Afterwards, we provide a partially centralized approach that reduces, at remarkable levels, fronthaul overheads, offering a flexible and scalable solution in large-scale antennas C-RAN. The proposed approach can also be adopted in H-CRANs, and H-CRAN related issues are also discussed.



\section{ System Architecture and Challenging Issues } \label{Background}

 \subsection{Heterogeneous Cloud Radio Access Network with Large-scale Antennas} 

In C-RANs, BBUs, which process radio signals, are separated from cell sites and concentrated at a data center (BBU pool), whereas in conventional RANs a BBU and a radio unit are located together. The cloud base station structure allows more advanced techniques for the management of radio resources, interference, computing resources, and energy consumption. 
In addition, due to the centralized infrastructure, mobile operators are able to significantly reduce the installation and operation cost.


 In H-CRANs, high-power macro cells with a large number of antennas and low-power small cells with fewer antennas can be combined to enhance network capacity and access network connectivity.
The performance of each user particularly in hot-spots or indoor buildings can be highly improved thanks to abundant reuse of limited spectrum resources and better link qualities. Meanwhile, mobile users can be adaptively or simultaneously served by small cells and macro cells to achieve high data rates and to maintain connectivity with reduced control-plane signaling for handover.
 
  As large-scale antennas are deployed in H-CRANs, the potential of H-CRAN can further be maximized with spatial multiplexing and array gain. At hotspots, many users can be spatially multiplexed to further increase per-user throughput; even cell-edge users who require high data rates can be satisfied by large-scale antenna beamforming of macro cells or by CoMP between macro cells and near small cells. 
  Abundant spatial degree of freedom provides a chance to handle cross-tier or co-tier interference in  complicated H-CRANs, avoiding underutilization of expensive spectrum resources.
 

 In H-CRAN design, conventional or already deployed macro base stations, termed eNB in LTE, which perform the entire baseband processing and upper layer functions can be combined with newly deployed RRHs that work for small and macro cells as well.
 Such macro base stations can be connected to C-RANs through X2 interface for inter-cell coordination in handover and interference management. With standalone baseband processing and non-ideal backhaul, however, fully centralized coordination with large-scale antennas is a challenging issue, due to a fundamental difficulty in CSI gathering\cite{Mugen_WCMAG_2014}. On the other hand, designing H-CRAN with only RRHs provides chances of flexible cell virtualization and computing resource pooling, as well as PHY-layer multi-cell coordination. 

  
 Suitable for H-CRAN deployment are wavelength-division multiplexing (WDM) and optical transport network (OTN) solutions, especially when fiber resources are limited and fronthaul transport network topology is complex\cite{WDM_COMMAG_2013}. 
 The bandwidth of the fronthaul link can be improved greatly as 40 to 80 optical wavelengths can be multiplexed in a single optical fiber.
 
 Considering that a tolerable round trip time of 400 usec for LTE-A (Long Term Evolution-Advanced) and a propagation delay of 5nsec/m in the optical fiber, a maximum distance of 40 km can be supported between a BBU pool and RRHs. This allows hundreds or even thousands of RRHs including macro-RRHs, to be connected to one BBU pool. In such a fronthaul transport network, numerous RRHs should share some fiber resources with other RRHs, limiting the end-to-end fronthaul resource for each RRH. 
 The system architecture of H-CRAN is depicted in  Fig. \ref{fig:Fig_Scenario_HCRAN}.

 \subsection{Challenging Issues in Large-scale Antenna Operation in H-CRAN} 

\emph{\bf{Joint MIMO Processing and User Scheduling}}:
High-power macro-RRH may cover tens/hundreds of low-power small-RRHs in H-CRANs. Coverage-different and numerous interference-connected cells in H-CRANs make RRH clustering and joint precoding/decoding more complicated than C-RAN. When a macro-RRH is equipped with many more antennas, the radio signals of macro-RRH can be generated to eliminate or mitigate interferences to small-RRH users while serving macro-RRH users with abundant degree of freedom. However, if  the degree of freedom of a macro-RRH is not relatively large compared with the number of users served by small-RRHs, there should be performed tightly-coupled precoding and scheduling over the macro-RRH and small-RRHs. The number of active antennas at macro- and small-RRHs can be adaptive to the number of small-RRH and macro- and small-RRH users.


\emph{\bf{CSI Gathering and Pilot Resource Management}}:
To obtain huge CSI with reasonable pilot overheads, a promising option is TDD-based large-scale antenna systems~\cite{Rusek_SPMag13}. In TDD-based H-CRAN, the received pilot signals of macro- and small-RRHs can be jointly processed at the BBU pool. To avoid correlated interference by pilot contamination, which limits the performance of massive MIMO even with an infinite number of antennas, integrated pilot resource management and channel estimation technique needs to be investigated. The main considerations are asymmetric coverage and the number of associated users of macro- and small-RRHs as well as the applied precoding schemes and resource sharing/partitioning among macro/small-RRHs.



\emph{\bf{Adaptive Cell Configuration}}:
In an H-CRAN consisting of high power macro-RRHs and low power small-RRHs, the entire network capacity and per user performance can be enhanced by exploiting multi-RRH joint transmission/reception with flexible cell configuration as in \cite{FluidNet_mobicom_2013,Mugen_WCL_2014}. To perform cooperative MIMO processing and resource allocation, RRHs should be adaptively clustered and occupy the necessary computing resource in a BBU pool. However, the required computing resources of each cell dynamically varies in H-CRAN environments due to various patterns of cell coverage and traffic loading. The BBU pool should assign computing resources to RRHs considering their baseband processing complexity. For real-time baseband processing and efficient computing resource use, an accurate amount of required computing resources should be modeled for large-scale antenna H-CRANs.







\emph{\bf{Fronthaul Resource Management and Topology Optimization}}: In complicated heterogeneous networks, where many RRHs are multi-hop connected to the BBU pool, a significant problem that must be solved is fronthaul resource managements, especially for fronthaul resources shared by hot-spot RRHs. In addition, fronthaul transport network topology and fiber resource installation should be decided considering each cell's user traffic and fronthaul data volume.
Resource and topology should be jointly optimized adaptive to heterogeneous number of antennas, coverage, and traffic or on/off state of each RRHs. Depending on the fronthaul resource medium, i.e., wireless channels or wavelengths in optical fiber, different approaches are needed for resource management and topology optimization.

\emph{\bf{Fronthauling and Architecture Design}}:
 Explosive fronthaul data volume is a very challenging issue, especially in H-CRANs, consisting of macro-RRHs, which are probably equipped with a large number of antennas, and small-RRHs, which are equipped with fewer antennas but much more densely deployed than macro-RRHs. The fronthaul data explosively increases as wireless networks are evolved with scaling the number of antennas, cell density, and spectrum expansion.
What are needed to tackle such explosively increasing fronthaul overheads in H-CRANs, are efficient and flexible fronthauling and H-CRAN architecture. 


\section{Fully Centralized C-RAN and Fronthaul Transport Solutions}
    \begin{figure*}
    \centering
    \includegraphics[width = 7in]{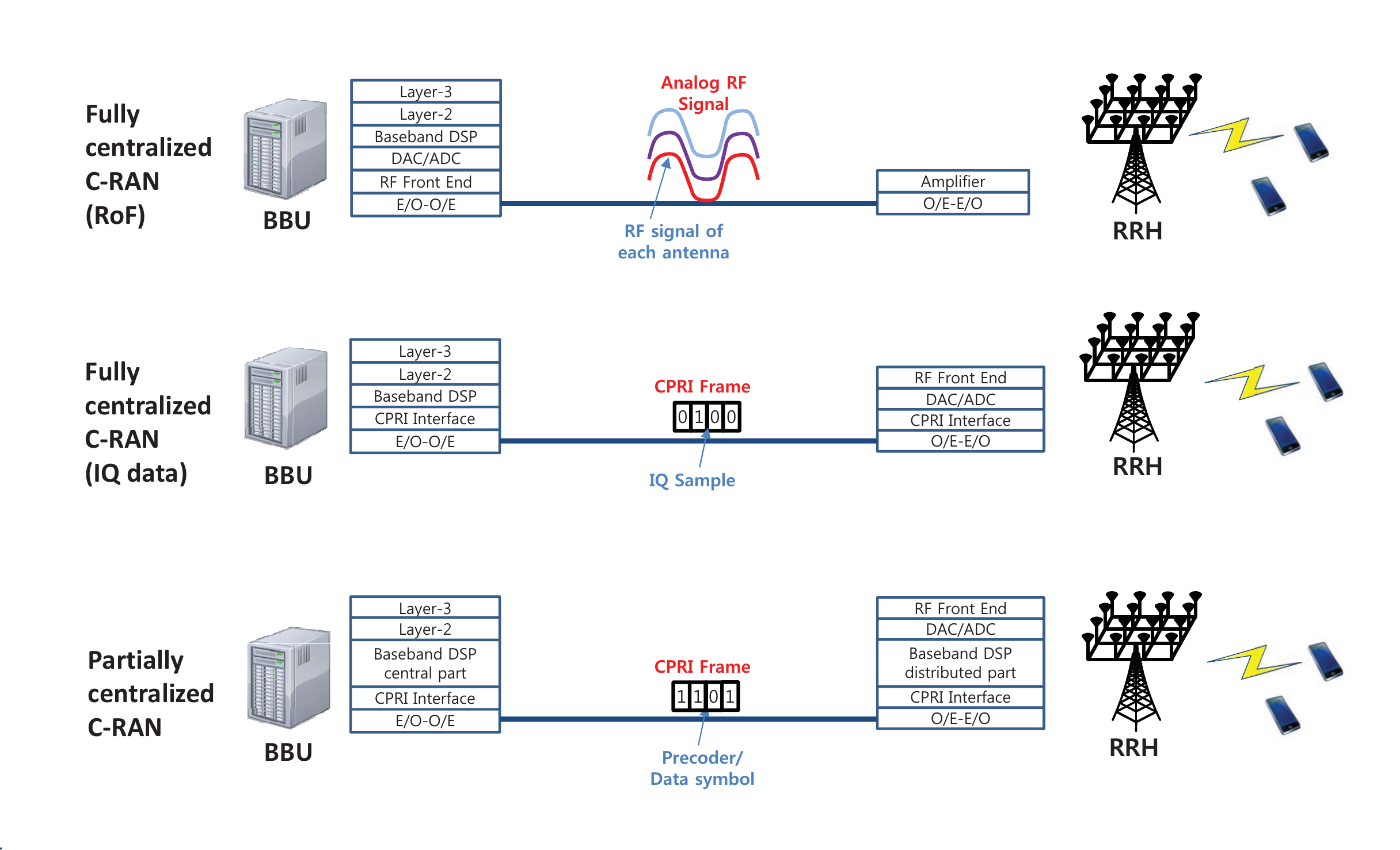}
    \caption{Fronthaul transport and C-RAN architecture solutions. } \label{fig:Fig_Fronthaul_comparison}
    \end{figure*}
	In this section, we describe existing fully centralized C-RAN (FC-RAN) solutions;
	one is radio over fiber (RoF), which carries radio signals directly over the optical fiber, and the other is digitized IQ data transport, which carries digitized IQ sample data over it.

 \subsection{Radio over Fiber}  
 
A way of communication between the BBU pool and RRHs is to employ radio frequency signal transmission based on RoF techniques \cite{FluidNet_mobicom_2013,RoFCRAN_JLT_2013}.
 In RoF systems, radio signals are carried in optical form between the BBU pool and RRHs before being radiated over the air.
 
In the BBU pool, digital baseband signals generated by a digital signal processor (DSP) are converted to analog radio frequency signals via a digital-analog converter (DAC) function. Then, the RF signal is carried over an optical signal using electro-optic modulators (EOM) such as Mach-Zehnder modulators and electro-absorption modulators.
 As radio frequency signals are completely manipulated in the BBU pool, the design of RRHs can be further simplified. 
Due to its transparency with respect to radio access technologies, RRHs can transmit (or receive) any radio signal (LTE, 3G, WiFi, etc) that can be transported over a fiber cable. By adopting WDM in fronthaul link, a set of radio signals for multiple RRHs\cite{RoFCRAN_JLT_2013} or multiple antennas can be multiplexed in a single fiber. 
 
Due to high carrier frequency and large frequency gap between wavelength channels (12.5-100 GHz), one wavelength can carry an antenna's radio signal nearly regardless of its bandwidth. With regard to bandwidth expansion in future cellular systems, this is an attractive prospect. However, each wavelength carries the radio signal of only one antenna at a time.
 In implementing H-CRANs, carrying radio frequency signals of large-scale antenna macro-RRHs and numerous small-RRHs requires huge wavelengths and fibers in the shared fronthaul links between RRHs and a BBU pool. Therefore, adaptive wavelength resource management and limiting RF-chains with proper large-scale antenna techniques are essential for RoF based large-scale antenna H-CRAN deployment.

 \subsection{Digitized IQ Data Transport}

 In conventional FC-RAN, the BBU pool and RRHs exchange baseband signal in the form of digitized IQ samples. The IQ samples are encapsulated using a fronthaul transport interface such as CPRI.
 In contrast to the RoF transport solution, digitized IQ sample information can be transported over both wired and wireless fronthaul links, resulting in easier implementation of small cells in H-CRANs.

The BBU generates a baseband signal for each antenna in the form of a digital IQ sample using a digital signal processing function after upper-layer protocol processing.
 As a result, the required fronthaul bit rate is given by 
 \begin{equation}\label{eq:R_IQ}
  R_{{\rm{IQ}}}  = \alpha Mb_{{\rm{IQ}}} f_{\rm{s}} 
 \end{equation} 
 where $b_{\text{IQ}}$ is the number of bits to represent a pair of IQ samples and $\alpha$ is the redundancy in the fronthaul transport interface.
 In CPRI, $\alpha  = \left( {\frac{{10}}{8}} \right)\left( {\frac{{16}}{{15}}} \right) = \frac{4}{3}$ considering on 8B/10B line encoding and the control channel portion of 1/16 \cite{CPRI}. The IQ sampling rate, $f_s$, is proportional to the wireless system bandwidth. This representation holds for both downlink and uplink. 
 Assuming a 20-MHz LTE system ($f_s$ = 30.72 MHz) with 64 RRH-antennas and 15 bits of IQ sample width ($b_{{\rm{IQ}}}$ = 30), one end-to-end CPRI link should be able to support 78.64 Gbps of IQ data. The current maximum rate of a CPRI link is around 10 Gbps\cite{CPRI}.
 
 In H-CRANs where many macro-RRHs and small-RRHs are equipped with many antennas, network-wide fronthaul data drastically increases as cell density increases or antennas are scaled up.
 To this end, some options are available to reduce the fronthaul data volume for a given number of RRH-antennas and wireless bandwidth.

\subsubsection{IQ data compression}
With compression/decompression at the BBU pool and RRHs, the IQ-sample data can be reduced at the cost of distortion in reconstruction of baseband signals and additional complexity both in the BBU pool and RRHs. In \cite{SurveyCRAN_COMST_2014}, the performance of existing IQ compression schemes are summarized. A joint compression/decompression scheme \cite{Compression_SPM_2014} exploiting correlation among the signals of neighbor RRHs might be extensible to H-CRANs. 

\subsubsection {Advanced MIMO technique}
 To reduce the number of active antennas, only the selected set of antennas can be used for radio signal transmission, as in \cite{LAU_TSP_2014}. Or limited RF-chains can be used with two-stage precoding techniques such as those in \cite{RFprecoding_TSP_2014}\cite{Hybrid_WCL_2014}, with which long-term RF-precoding for antenna elements is performed at RRHs and short-term baseband-precoding for RF-chains is performed at the BBU pool.
 With such schemes, the benefits of large-scale antennas can be exploited with reasonable fronthaul overheads. 

\section{Partially Centralized C-RAN for Large-Scale Antenna Operation}
 
  \begin{figure*}
  \centering 
      \subfigure[Downlink structure]{\label{subfig:Fig_Structure_DL}
              \includegraphics[width=7in, height = 2in]{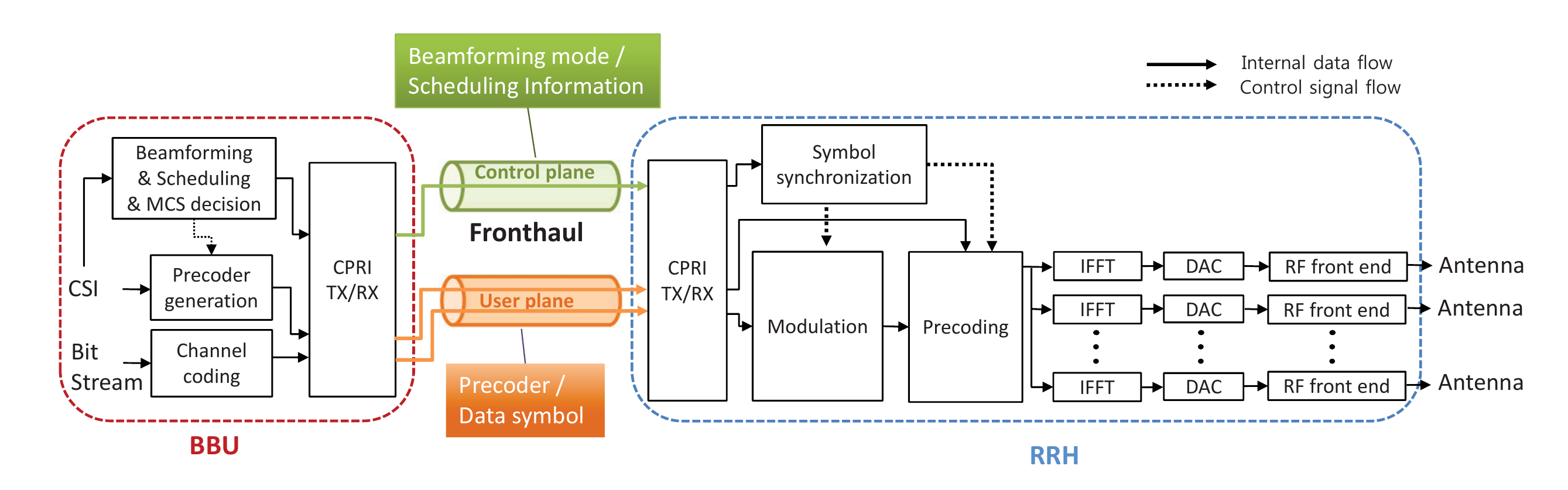}}
      \subfigure[Uplink structure]{\label{subfig:Fig_Structure_UL}
              \includegraphics[width=7in, height = 2in]{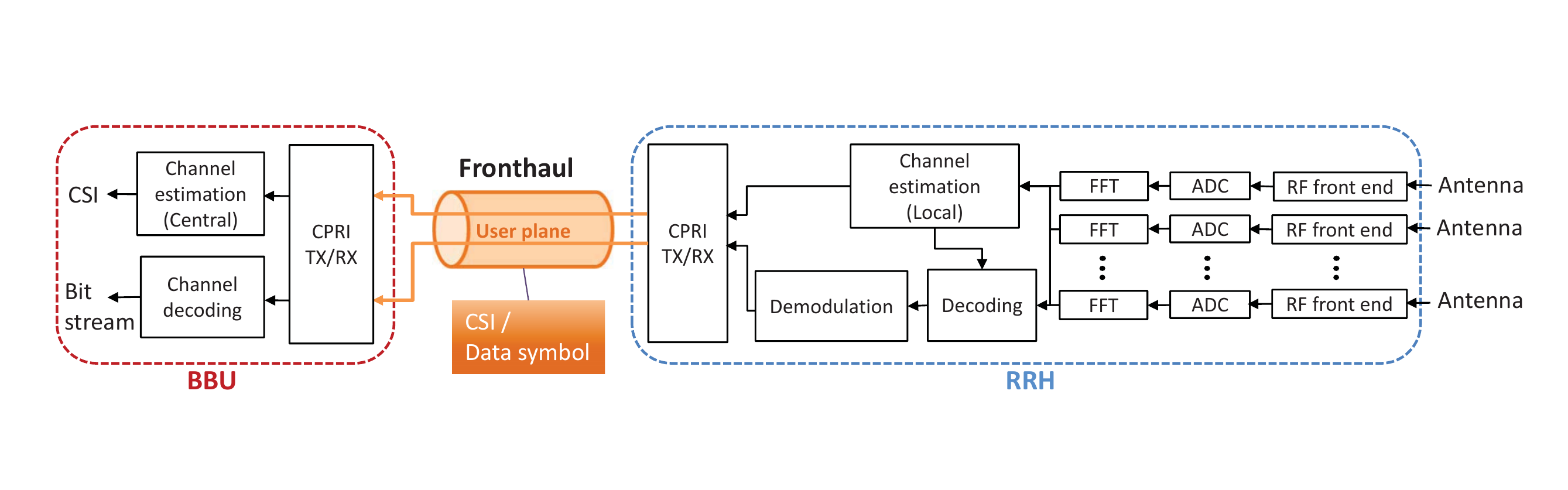}}
                       
  \caption{Example of the partially-centralized C-RAN (PC-RAN) structure for large-scale antenna operation.} \label{fig:Partial_CRAN_Architecture}
  \end{figure*}
      
 The fundamental limitation of FC-RAN is that per-antenna baseband (or RF signal in RoF based C-RAN) information should be carried in fronthaul, resulting in huge requirements of fronthaul resources with order-of-antenna numbers when large-scale antenna systems are the case.
 As an alternative, partially centralized solution \cite{CMRI}\cite{SurveyCRAN_COMST_2014} can be adopted where L1 (PHY-layer) processing is co-located with RRHs to avoid fronthaul-inefficient IQ-sample transport. However, processing all L1 functions at RRHs as the original concept in \cite{CMRI} sacrifices the joint L1/L2 cooperation gain among multiple RRHs. 
 One desirable solution for reducing fronthaul data but maintaining cooperation gain is dividing PHY-layer functionalities into the central part and the distributed part. Fig.~\ref{fig:Fig_Fronthaul_comparison} shows the key differences between the FC- and our proposed PC-RAN.

 \subsection{Basic Concept}
Fig. \ref{fig:Partial_CRAN_Architecture} presents a PC-RAN architecture for TDD based large-scale antenna operations.
One of the fundamental features of the proposed C-RAN is that the baseband processing part is divided into the BBU part and the RRH part.
By doing so, precoder, data symbols, and channel information are separately transported instead of heavy IQ data. 
  
The BBU pool may jointly decide the precoder and scheduled users with proper modulation and coding scheme (MCS) levels considering their channel conditions.
The main products of the BBU processing are modulated data symbols and precoders (decoders for uplink).
Precoding vectors and data symbols for the scheduled users are transported to corresponding RRHs which transmit radio signals through their antennas after precoding received data symbols. 

 \subsubsection{Data symbol transport}
 
 In PC-RAN, by transporting constellation points (after deciding the modulation scheme), a data symbol can be represented with a smaller number of bits compared with a quantized IQ sample.
 Moreover, the data volume size can be reduced to the order of the number of users ($K \ll M$) from the number of antennas. Considering the pilot overhead for channel estimation in a slot, the required bit rate for data symbol delivery of $K$ users in TDD systems is
 \begin{equation}\label{eq:R_DS}
    R_{{\rm{DS}}}  = \alpha \left(1  - \frac{\tau }{T}\right)K b_{{\rm{DS}}} f_{{\rm{sym}}},
 \end{equation}
where $f_{{\rm{sym}}}$ is the symbol rate. In 20 MHz OFDM systems with 1200 subcarriers ($N_{sc}$ = 1200), $f_{\rm{sym}}$ =  $\frac{N_{sc} }{T_s + T_g}$ = 16.8 MHz. 
 $\frac{\tau }{T}$ is the ratio of the number of symbols used for the uplink pilot, $\tau$, to the number of total symbols, $T$, in a transmission slot. 
 The number of bits to represent one data symbol, $b_{{\rm{DS}}}$, depends on the representation method as described below.
 
  \subsubsection{Precoder transport}
   
 For consecutive symbols during a coherence time (or slot time during which the channel condition is assumed to remain constant), the same precoder can be applied. Even though the length of the precoding vector increases with the number of RRH antennas, the precoding data volume is bearable due to its low transport frequency compared to data symbols. 
 The required bit rate for a precoder update to support $K$ users (or streams) with $M$ antennas is 
  \begin{equation}\label{eq:R_Pre}
    R_{{\rm{Pre}}}  = \alpha MKb_{{\rm{IQ}}} f_{{\rm{Pre}}}.
  \end{equation}
 The frequency of precoder update, $f_{{\rm{Pre}}}$, can be set as $(\frac{1}{T})f_{{\rm{sym}}}$ according to the transmission slot of the wireless physical channel.

 \subsection{Centralized and Distributed Precodings}
    \begin{figure*}
    \centering 
        \subfigure[Centralized and distributed precodings for a single RRH]{\label{subfig:Fig_Single_RRH_CPR_DPR_scenario}
                \includegraphics[width=3.3in, height = 3in]{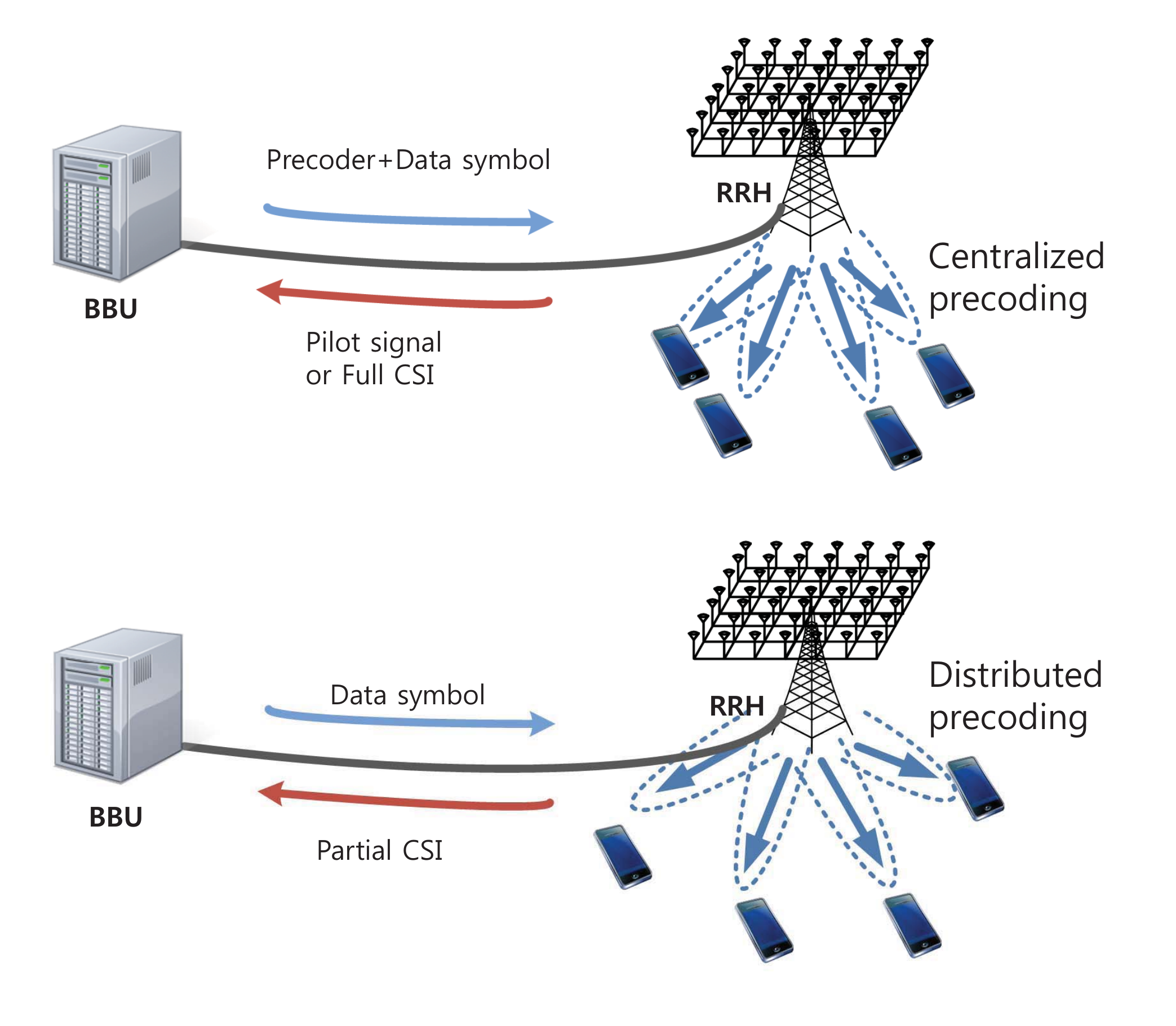}}              
        \subfigure[Centralized/distributed precoding scenario in H-CRAN]{\label{subfig:Fig_HCRAN_CPR_DPR_scenario}
                \includegraphics[width=3.3in, height = 3in]{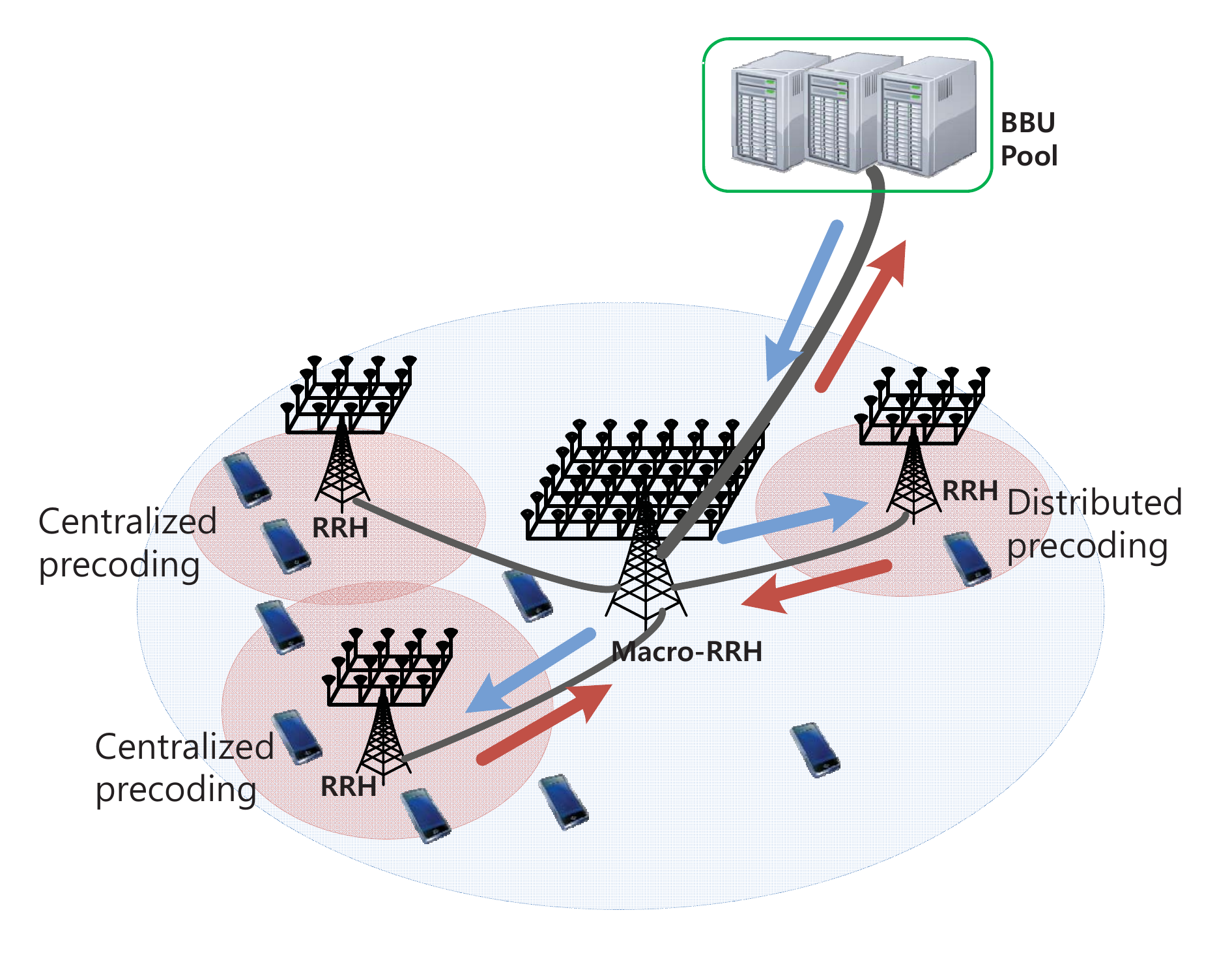}}
                         
    \caption{Centralized and distributed precoding scenarios. In  ultra-dense H-CRAN, distributed precoding can be exploited to reduce fronthaul burden according to RRH locations and user channel conditions.} \label{fig:CBF_DBF_scenario}
    \end{figure*}
 Multiple antenna precoding schemes have different characteristics with respect to the multiplexing order, user channel environments, and algorithm complexity. In the proposed PC-RAN, precoder design can, as shown in Fig. \ref{subfig:Fig_Single_RRH_CPR_DPR_scenario}, be done in two ways; {\it centralized} and {\it distributed} precodings according to the need for centralized cooperative processing over multiple RRHs, the computational burden on the RRH, and available fronthaul bandwidth.

\subsubsection{Centralized precoding}	
 
  To overcome cross-tier and co-tier interference in densely deployed H-CRANs, the BBU pool may perform a cross-tier and/or co-tier cooperative processing over multiple RRHs, aiming at network-wide performance optimization. For some users or RRHs that experience severe interference, it would be desirable to use zero-forcing based multiuser MIMO or network MIMO schemes. Large-scale zero-forcing based precoding for a macro-RRH and multiple small-RRHs should be performed in a BBU pool where the required computing resources can be flexibly assigned and CSI of entire RRHs are gathered.
  
  To perform centralized precoding, the following information should be exchanged between a BBU pool and RRHs.
 
 $\bullet$ Data symbol: modulation scheme and bit symbols for scheduled users. 
 
 $\bullet$	Precoding vector: precoding weight vectors to be precoded with data symbols of spatially multiplexed users. 
 
 $\bullet$ Estimated full CSI or pilot signal: information needed to decide the precoder in a BBU pool.
 
There are two ways to deliver the CSI of each user to the BBU: 1) The RRH transports the locally estimated CSIs (LE-CSIs) of selected users to the BBU, and 2) the RRH transports the received pilot signal during the channel training period in the form of an IQ sample, as in conventional C-RANs. The second option is needed when the BBU pool performs cooperative MIMO processing and channel estimation over multiple RRHs to eliminate inter-cell interference or pilot contamination\cite{Yin_JSAC_2013}.

 \subsubsection{Distributed precoding}	
 For distributed precoding, RRHs generate precoding vectors for scheduled users using LE-CSIs during the channel training period. One representative application of distributed precoding is conjugate beamforming, which aims to maximize the desired signal power regardless of interference and requires low complexity to obtain the precoder \cite{Argos_mobicom_2012}.  
 Zero-forcing based precoding also can be used if the computational complexity is supportable with not so many antennas and spatially multiplexed users. Non-zero-forcing based limited coordinations among RRHs are also possible in distributed precoding. 
 The following information should be exchanged between a BBU pool and RRHs.
 
 $\bullet$	Data symbol stream: Modulation scheme and bit symbols for scheduled users. 
 
 $\bullet$	Precoding vector: Not necessary. The precoding vector can be locally calculated at each RRH using LE-CSIs if the precoding technique and scheduling information is informed by the BBU.
 
 $\bullet$	Partial channel information: For distributed precoding operation, there is no need for the BBU to know instantaneously the full CSI of users because RRHs calculate the precoder for each user using LE-CSI. 
 

The separation of precoder and data symbol in fronthaul is more beneficial in the distributed precoding scenario where precoding weights not transported over fronthaul and the information about CSI or training signal can be further reduced. 
However, the BBU may need to know the channel condition to decide the appropriate precoding technique and to schedule each user. Especially, for the MCS decision, the BBU has to receive the expected SINR of considered users from RRHs or infer it from the effective channel gain of the users and long-term channel gains between the users and neighbor RRHs.
 
          \begin{figure*}
          \centering            
        
             \subfigure[Aggregate RRH sum-rates]{\label{subfig:Fig_Graph_sumrate_HCRAN_macro_256_64_small_64_32}
                      \includegraphics[width=3.4in, height = 2.9in]{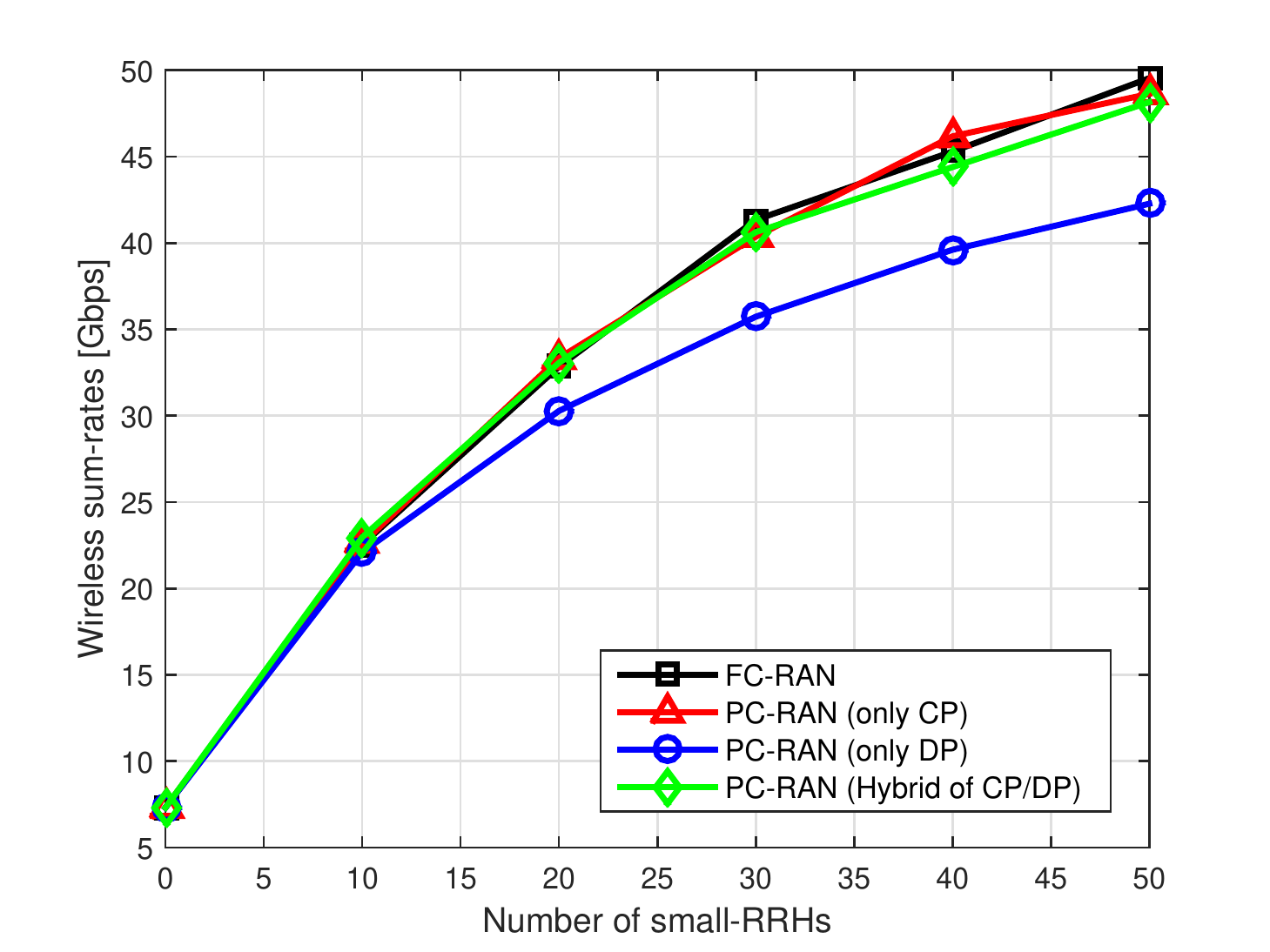}}   
             \subfigure[Aggregate required fronthaul rates]{\label{subfig:Fig_Graph_fronthaulrate_HCRAN_macro_256_64_small_64_32}
                      \includegraphics[width=3.4in, height = 2.9in]{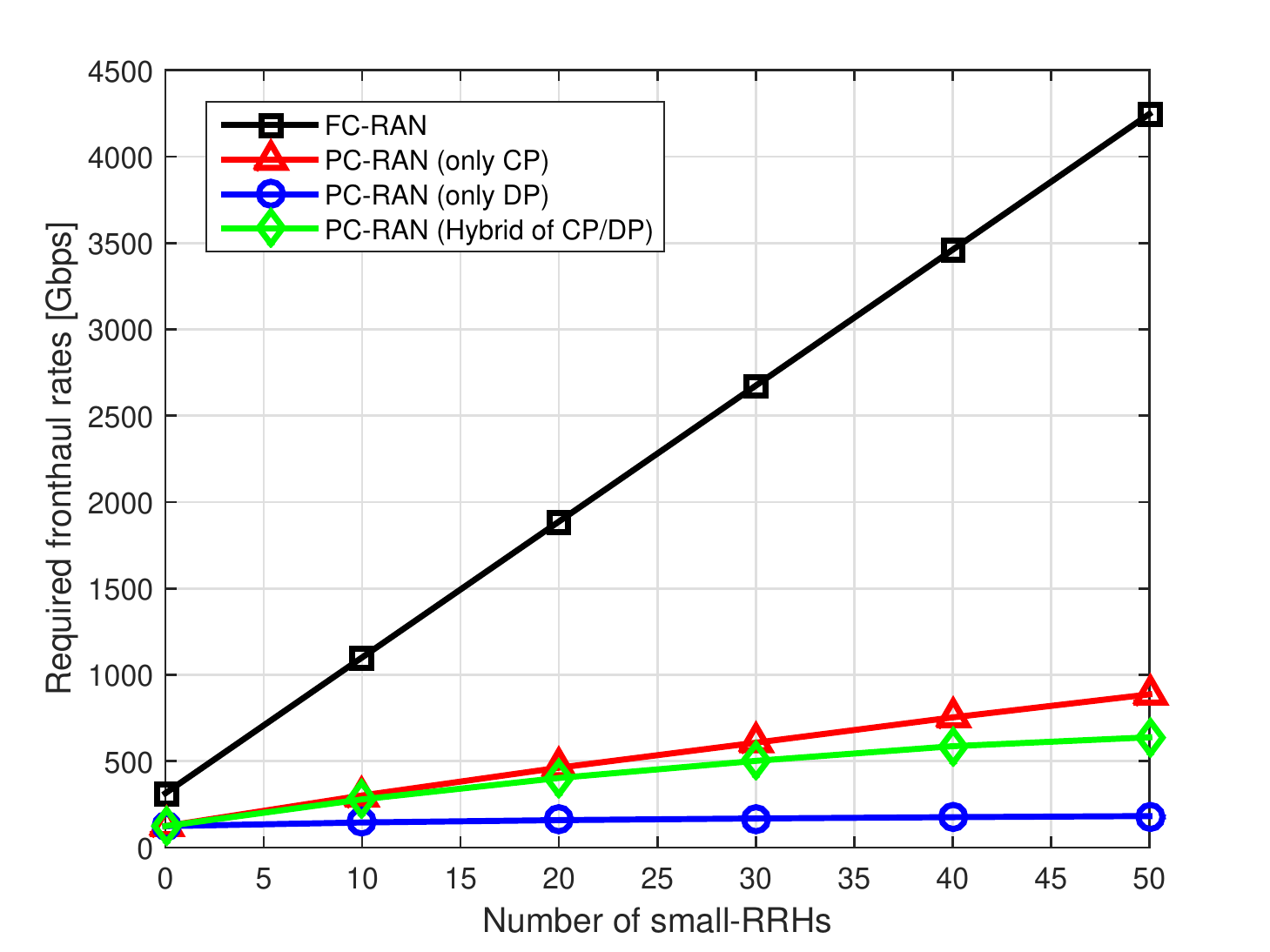}}   
       
          \caption{Wireless sum-rates and required fronthaul rates in an H-CRAN according to the number of small-RRHs in a macro-RRH coverage. 500 users are randomly deployed in a radius of macro cell coverage, 2 km. Macro- and small-RRHs are equipped with 256 and 64 antennas, respectively. } \label{fig:Fig_Graph_performance_in_HCRAN}
          \end{figure*}

 \subsection{CSI Estimation and Report}	
The TDD operation of large-scale antenna systems does not need pilot resources proportional to the antenna array size 
because uplink and downlink channels are estimated  
by the channel reciprocity and pilot signals sent by users. 
 In such TDD operation scenarios, the RRH can estimate the channel of each user if it knows the location of pilot resources assigned to each user. For such localized channel estimation, the BBU should inform the RRH of the pilot resource map of users served by the BBU.
 The required uplink fronthaul bit rate to report locally-estimated CSI (LE-CSI) is
 $R_{{\rm{LE-CSI}}}  = \alpha MKb_{{\rm{IQ}}} f_{{\rm{CSI}}}.$
 The frequency of CSI report for each user, $f_{{\rm{CSI}}}$, is given as $(\frac{1}{T})f_{{\rm{sym}}} $ (in this case, $R_{{\rm{LE-CSI}}}=R_{{\rm{Pre}}}$ ) and it can be adaptively chosen by balancing CSI accuracy and the fronthaul resource budget. 
 
 Another option is to transport IQ samples of received pilot signals to the BBU pool, 
 especially when cooperative channel estimation against pilot contamination or for CoMP is needed. In such a case, the required uplink bit rate is    
  $R_{{\rm{pilot,IQ}}}  = \alpha M(\frac{\tau }{T})b_{{\rm{IQ}}} f_{{\rm{sym}}} $. 
  
 
 When only partial channel information, such as expected SINR or effective channel gain of users, is needed for distributed precoding, the required uplink bit rate to report the partial channel information obtained using LE-CSI is $ R_{{\rm{LE-CSI,partial}}}  = \alpha Kb_{{\rm{real}}} f_{{\rm{CSI}}}  $, where $b_{{\rm{real}}}$ is the number of bits needed to represent one real value.

  \subsection{Operation of Centralized/Distributed Precoding in H-CRAN}
  In PC-RAN, the required fronthaul bandwidth for centralized precoding is much greater than that for distributed precoding because the required bit rates for the precoder and pilot signal IQ data outweigh that of data symbols and increase with the number of antennas. On the other hand, the fronthaul data volume of distributed precoding operations does not depend on the antenna array size because it is unnecessary to exchange precoder and full CSI (or pilot signal IQ data) between a BBU pool and an RRH. In this sense, the hybrid operation of centralized and distributed precoding provides chances to flexibly utilize fronthaul resources. In H-CRANs, either centralized or distributed precoding can be applied for each RRH depending on the existence of possible interference victim users. For example, a centralized precoder, as depicted in Fig. \ref{subfig:Fig_HCRAN_CPR_DPR_scenario}, can be applied only for small-RRHs that may interfere significantly with macro- or other small-RRH users to minimize fronthaul traffic overhead.

%
%
%
  
  Fig. \ref{fig:Fig_Graph_performance_in_HCRAN} shows the aggregated wireless sum-rates and fronthaul traffic in an H-CRAN consisting of one macro-RRH and many small-RRHs. We can see in the figure that the required fronthaul traffic is remarkably reduced when the H-CRAN is implemented with PC-RAN compared to FC-RAN. More interestingly, with the proper hybrid operation of centralized/distributed precoding for small-RRHs, fronthaul traffic is further reduced. At the same time, the wireless sum-rate is somewhat degraded due to uncontrolled interference but is still much better than when only distributed precoding is applied. The main challenge is to find a good compromise between cell/network throughput and fronthaul overhead. 
  It is also worth mentioning that important considerations in H-CRAN design are the number of macro- and small-RRH antennas and RRH density, which impact the entire capacity and required fronthaul capacity of the H-CRAN.

     \begin{table*}
     \centering
     \caption{Characteristics of C-RAN architecture - fronthaul transport solutions  }
     \label{Table_fronthaul_solution}
     \begin{tabular}{|c|c|c|c|c|c|c|c|}
     \hline
     ~~~~           & Signal form  & Joint RRH   & Spectrum  &  Antenna  & RRH  &  Possible \\
     &  in fronthaul &  processing  &  expansion &  scalability  & complexity &  medium \\     
          
     \hline
     \hline
     FC-RAN  & Analog RF & Possible  & Almost  & Bad & Very low & Wired  \\
      (RoF)  & &   & unlimited & & & (fiber)  \\
     \hline
     FC-RAN   & Digital & Possible   & Limited & Bad & Low & Wired \\
      (IQ data)  & baseband &  & &  & & /Wireless \\
     \hline    
     PC-RAN   & Digital & Possible    & Limited & Good & Normal & Wired \\
     (centralized  & baseband &  & & & & /Wireless \\     
      precoding)  & &  & &  & & \\ 
     \hline    
     PC-RAN   & Digital & Limited    & Limited & Good & Normal & Wired\\
     (distributed &  baseband &  & &  &  &/Wireless  \\          
       precoding)  & &  & &  & &  \\          
          
     \hline    
    
     \end{tabular}
     \end{table*}

\section{Discussion}

    As investigated in the previous section, the partially centralized C-RAN approach provides chances to significantly reduce fronthaul traffic or flexibly balance between the wireless performance and fronthaul overhead. For such a flexible operation, however, RRHs should be more complicated than FC-RANs to perform local channel estimation and precoding while tightly synchronized with their BBU pool. The features of the addressed C-RAN architectures and corresponding fronthauling solutions are summarized in Table \ref{Table_fronthaul_solution}. Finally, we remark on the following related issues of our proposed PC-RAN.

 \emph{Adaptive CSI Gathering}: 
 For large-scale antenna operation, CSI or pilot signal transport requires a large amount of fronthaul resources. In PC-RAN, the CSI traffic can be remarkably reduced according to the precoding scenario (i.e., centralized or distributed) and the target accuracy of CSI values thanks to local CSI estimation at RRHs. For example, in the case of centralized precoding, RRHs can report locally estimated CSIs less frequently than pilot signal transmissions at the expense of some precoder performance degradation. On the other hand, for distributed precoding, RRHs do not need to report full CSIs. However, it is possible that RRHs report part of locally estimated CSIs for more accurate interference estimation or for neighbor RRHs that may perform centralized precoding. Thus, CSI reporting of RRHs can be adaptively controlled by balancing between fronthaul overhead and performance gain of edge-users. This is also one of the key merits of applying PC-RAN, but it is quite challenging to consider the relation between precoding, CoMP communication, and channel estimation.    


 \emph{Fronthaul Resource Constrained Precoding}: 
 The required fronthaul bit rates depend on many factors including where the precoding is calculated, the number of active antennas (or RF-chains) at RRHs, user multiplexing order, uplink sounding period, precoder and scheduling granularity. Intuitively, by jointly deciding a proper precoding scheme and user scheduling with a consideration on the given fronthaul conditions and required wireless performances, limited fronthaul resources can be more efficiently utilized \cite{SKPARK_JCN_2013}.
  In H-CRANs, as numerous macro- and small-RRHs are connected to a shared fronthaul link to a BBU pool, available fronthaul resources for each RRH can be limited when the network is heavily loaded. In such a case, the number of active antennas  and multiplexed users can be optimized considering the number of associated users in each RRH and cross-tier interference.

  \emph{Joint Processing in Uplink}: In downlink, data symbols and precoders of coordinating RRHs are simultaneously decided in the BBU pool, and generated data symbols and precoders are transported to corresponding RRHs. In uplink, however, signals from coordinating RRHs should be collected in the BBU pool and jointly detected to achieve joint processing gain to the fullest. Thus, it is difficult to exploit the separation of data symbol and precoder of PC-RAN when uplink multi-RRH cooperation is needed.

 \emph{Frequency Division Duplexing Operation}: The concept of precoder and data symbol separation holds for FDD systems. In FDD systems, however, downlink CSIs are not available at RRHs because CSI feedback from user terminals is transparent and indistinguishable to RRHs. Thus, in FDD systems, it is difficult to exploit distributed precoding and the CSI-related fronthaul burden due to the fact that large-scale antennas can not be relieved.

  \emph{Advanced precoding techniques for large-scale antennas}: If a two-stage baseband/RF precoding scheme \cite{RFprecoding_TSP_2014}\cite{Hybrid_WCL_2014} is adopted in FC-RAN, the antenna-scale dependent fronthaul data volume can be reduced proportional to the number of RF-chains, $S$. The two-stage can also be applied in the proposed PC-RAN. Then, the size of precoder and CSI information can also be reduced.  
When $f_{{\rm{Pre,BB}}}$ and $f_{{\rm{CSI,BB}}}$ are update frequency of baseband precoding and effective CSI for baseband precoding, and $f_{{\rm{Pre,RF}}}$ and $f_{{\rm{CSI,RF}}}$ are for RF precoding, the required fronthaul rates for precoder and CSI are
$R_{{\rm{Pre}}}  = \alpha ( SK b_{{\rm{IQ}}} f_{{\rm{Pre,BB}}} + MK b_{{\rm{IQ}}} f_{{\rm{Pre,RF}}})$ and 
$R_{{\rm{CSI}}}  = \alpha ( SK b_{{\rm{IQ}}} f_{{\rm{CSI,BB}}} + MK b_{{\rm{IQ}}} f_{{\rm{CSI,RF}}})$, respectively, in a centralized precoding operation.  For distributed precoding, the required fronthaul rates, while very low, are the same regardless of adoption of two-stage precoding technique.

\emph{Fronthaul data compression}:
IQ data compression/decompression schemes\cite{SurveyCRAN_COMST_2014} can reduce the fronthaul traffic of FC-RAN at the cost of some wireless performance degradation. Such IQ data compression schemes probably can also be applied to PC-RAN in a transporting precoder and CSIs. The impact of IQ data compression/decompression in PC-RAN still needs to be investigated. In PC-RAN, the number of bits for data symbol transport also can be reduced by grouping the data symbols based on applied modulation scheme instead of informing the constellation point of symbols.

\section{Conclusion} \label{Conclusion}

A promising solution to the problem of enhancing network capacity in cellular networks is deploying more small cells and installing a great number of antennas. Large-scale antenna operation on H-CRANs requires not just extreme processing resources but huge fronthaul data delivery. In this article, we explored existing C-RANs and promoted a partially centralized approach in H-CRANs with large-scale antennas. The partially centeralized C-RAN approach provides chances for mobile operators to efficiently utilize fronthaul resources and to flexibly adopt proper precoding or CoMP schemes according to available fronthaul resources and required wireless performance in H-CRANs.
 
%
%
%
%
%
%
%
%
 
%

\section*{Acknowledgment}
 This research was funded by the MSIP (Ministry of Science, ICT \& Future Planning), Korea in the ICT R\&D Program 2014.


\ifCLASSOPTIONcaptionsoff
  \newpage
\fi




%

\renewcommand{\baselinestretch}{1.0}
\bibliographystyle{IEEEtran}
\bibliography{references_WC_mag} 

%

\begin{IEEEbiography}{Sangkyu Park} received B.S degree in Electrical Engineering from Yonsei University, Seoul, Korea in 2009. He is currently working towards the Unified Master's and Doctor's Course in Seoul National University. His current research interests include heterogeneous network, cloud radio access network, and massive MIMO.
\end{IEEEbiography}

\begin{IEEEbiography}{Chan-Byoung Chae} is an Assistant Professor in the School of Integrated Technology, College of Engineering, Yonsei University, Korea. He was a Member of Technical Staff (Research Scientist) at Bell Laboratories, Alcatel-Lucent, Murray Hill, NJ, USA. He received his Ph.D. degree in Electrical and Computer Engineering from The University of Texas (UT), Austin, TX, USA in 2008, where he was a member of the Wireless Networking and Communications Group (WNCG). His current research interests include capacity analysis and interference management in energy-efficient wireless mobile networks and nano (molecular) communications.

\end{IEEEbiography}

\begin{IEEEbiography}{Saewoong Bahk} received B.S. and M.S. degrees in Electrical Engineering from Seoul National University in 1984 and 1986, respectively, and the Ph.D. degree from the University of Pennsylvania in 1991. From 1991 through 1994, he was with AT\&T Bell Laboratories as a Member of Technical Staff where he worked for AT\&T network management. In 1994, he joined the school of Electrical Engineering at Seoul National University and currently serves as a Professor. He has been serving as TPC members for various conferences including ICC, GLOBECOM, INFOCOM, PIMRC, WCNC, etc. He is on the editorial boards of IEEE Transaction on Wireless Communications (TWireless), Computer Networks Journal (COMNET), and Journal of Communications and Networks (JCN). His areas of interests include performance analysis of communication networks and network security. He is an IEEE Senior Member and a Member of Whos Who Professional in Science and Engineering.
\end{IEEEbiography}




\end{document}